\documentclass[pre,preprint,showpacs,preprintnumbers,amsmath,amssymb]{revtex4}
\usepackage{graphicx}% Include figure files

\usepackage{dcolumn}% Align table columns on decimal point
\usepackage{bm}% bold math
\usepackage[mathscr]{eucal}
\usepackage{mathrsfs}

\begin{document}

% Full title of the paper (Capitalized)
\title{State Variables and Constraints in Thermodynamics of Solids and Their Implications}

% Authors, for the paper (add full first names)
\author{Koun Shirai}
% \email{koun@sanken.osaka-u.ac.jp}
\affiliation{%
Nanoscience and Nanotechnology Center, SANKEN (ISIR), Osaka University, 8-1 Mihogaoka, Ibaraki, Osaka 567-0047, Japan
}%

\begin{abstract}
It is a common observation that the current properties of a solid depend on its past history, such as heat treatment. This fact causes a serious conflict with a principle of thermodynamics: a thermodynamic state is independent of the process by which the current state was obtained. 
The fundamental question is thus which are state variables for a solid that specify the current state without referring the past history.
The present study gives the answer to this question from the fundamentals of thermodynamics. A thermodynamic system is characterized by constraints, and one constraint corresponds to one state variable (thermodynamic coordinate, TC). Solids are characterized as having all the atoms constrained so as to occupy their own space distinguished from those of the other atoms. In equilibrium, the time-averaged position $\bar{\bf R}$ of an atom is unique and distinct from the others, which qualifies $\bar{\bf R}$ as a TC. Thus, the fundamental relation of equilibrium is specified by the set of all $\{ \bar{\bf R}_{j} \}$ together with the internal energy. The uniqueness and distinguishability of $\bar{\bf R}_{j}$ are essential for the TC, whereas the microscopic size of the quantity is irrelevant.
\end{abstract}

\pacs{}
% deflect from

\maketitle
%%%%%%%%%%%%%%%%%%%%%%%%%%%%%%%%%%%%%%%%%%
%% Only for the journal Gels: Please place the Experimental Section after the Conclusions

%%%%%%%%%%%%%%%%%%%%%%%%%%%%%%%%%%%%%%%%%%

\section{Introduction}
\label{sec:intro}
Thermodynamics is a universal theory, whose validity covers a wide range of physics from the scale of elemental particles to that of astronomy \cite{Gal-Or74,Fortov}. 
Although its fundamentals were established long ago, we still have conceptual difficulties with it. No textbooks explain what are the state variables of solids in a thermodynamic context \cite{thermo-texts}. 
For gas states, it is evident that there are only two independent variables, temperature $T$ and volume $V$ (or pressure $P$), when the number of particles $N$ is fixed. In a typical university curriculum, we first learn gas states in thermodynamics. Next, solid states are learned in statistical mechanics; we calculate the specific heat of solids on the basis of their microscopic structures. After that, there is no further study on thermodynamics. 
As a result, we hold the view that only $T$ and $V$ are independent state variables for solids too.
(Here, we consider only one-component systems with a fixed number of particles \cite{Note-anisotropy}). 
Then, we encounter severe difficulties. The magnetic properties of a ferromagnetic solid suffer influence of the past history of the applied magnetic field: the current properties cannot be uniquely determined solely by $T$ and $P$. This contradicts a fundamental of thermodynamics, that is, a thermodynamic state must be independent of the process by which the current state was obtained. Traditionally, this difficulty is circumvented by regarding hysteresis as an exceptional case and treating by nonequilibrium thermodynamics \cite{Bertotti05}. But, hysteresis is not a special property which only restricted materials carry. All materials exhibit hysteresis to a certain degree. It is known that the current properties of metals depend on the thermal treatments to which they were subjected. Every silicon wafer is not the same even at the same $T$ and $P$. Suppose that a silicon wafer is subjected to an electron bombardment; interstitial atoms and vacancies are created; then we see that the internal energy of the wafer is not the same as before even though the same $T$ and $P$.
From these observations, we see that the current state of any solid cannot be determined by the current values of state variables, as far as only $T$ and $P$ are regarded as the state variables for solids, leading to the absurd conclusion that thermodynamics does not hold for solids. 
% This threatens the position of thermodynamics as the first-ranked laws of physics.
Although a few people have addressed this problem \cite{Bridgman50, Kestin70,Maugin94}, most of the others do not consider them. %from the thermodynamic viewpoint. 

Confusion occurs when we are not careful for the distinction between state and process. It is a leaning in the first-course thermodynamics that the state of a system is independent of the process in which the current state of the system was obtained.
Consider a turbine generating electric power. A hot vapor flows into the turbine, with the temperature $T_{1}$ at the inlet. A wasted water flows out from the turbine, with the temperature $T_{2}$ at the outlet. Let us ask what is the value $T_{2}$ for a given value $T_{1}$. There is no unique answer. It depends on the process  throughout which the water is underwent. Inside the turbine, complicated exchanges in energy between the vapor and the blades occur. Heat loss due to friction with any part of the turbine occurs. The resulting properties of the rejected water depend on the process. In spite of this, the state of the water can be uniquely determined by $T$ and $P$, once thermodynamic equilibrium is recovered at the outlet. We do not need to know details of the process. Everybody understands this distinction for gases. However, when we investigate solids, everything is clouded, because we do not know state variables of solids, which uniquely specify the current state of a solid.

Thus, the essential question is which quantities are required to specify the states of a solids and how many variables are required. We learn in thermodynamics lessons that thermodynamic equilibrium states are characterized by a few state variables [3(b)-(e)]. However, what are state variables? We may answer that state variables are macroscopic properties characterizing the equilibrium. But, the formers have sense only in the latter. The argument turns to be circular. It is indeed difficult to give coherent definitions of state variables and equilibrium \cite{Tisza,Kline57,Hatsopoulos}. 
% evade or skip
Most textbooks avoid this problem and take these notions as {\it a priori} knowledge. This makes everything easy and allows us to benefit from all the advantages of thermodynamics. 
For a system composed of a large number of particles, the change in state can be described by the change in the internal energy $\Delta U$ with state variables $X_{j}$,
\begin{equation}
\Delta U = T dS + \sum_{j=1}^{M} F_{j} \cdot dX_{j},
\label{eq:first-law}
\end{equation}
without knowing details of the intermediate stages of the change. Here, $S$ is entropy and $F_{j}$ is a generalized force conjugate to $X_{j}$. In Eq.~(\ref{eq:first-law}), $M$ is the number of state variables other than $S$. The state space is thus spanned by $M+1$ state variables: here state variable is called {\it thermodynamic coordinate} (TC) after Zemansky and Dittman [3(b)]. Variables $X_{j}$ are mechanical part of TCs: deformable coordinates in Ref.~[3(f)]. Here, an obvious coordinate $S$ (or $T$) is taken as the zeroth coordinate of the state space, and we consider the remaining space of dimension $M$: an $M$-dimensional state space is understood in this meaning throughout this paper.
Although no textbooks impose a limitation on the dimensionality $M$, it is tacitly assumed to be small: the major advantage of thermodynamics is that the behavior of a large number of particles that constitute a macroscopic system can be described by only a few variables [3(c)]. To be worse, none of the textbooks gives concrete examples of TCs other than $T$, $V$, and $N$ (or their conjugates). 
However, suppose a long container of volume $V$ filled with a gas of $N$ molecules. The internal energy of the gas is described by three variables as $U=U(T,V,N)$. Let us next introduce a diathermal wall, which may or may not be mobile, in the container. Now, two volumes, $V_{1}$ and $V_{2}$, and two numbers of molecules, $N_{1}$ and $N_{2}$, are needed to describe the entire system as $U=U(T,V_{1},N_{1}; V_{2},N_{2})$ with conditions $V_{1}+V_{2}=V$ and $N_{1}+N_{2}=N$. This partitioning can be continued by adding as many diathermal walls as pleased. To say whether $M$ is small or large is nonsensical. 

The rigorous definition of TC is possible only from a principle more fundamental than thermodynamic equilibrium is: a fundamental notion cannot be explained by less fundamental laws. Only the second law of thermodynamics deserves to play the leading principle \cite{Gyftopoulos}. This study reveals what are TCs of solids from the fundamentals of thermodynamics.
The notion of TC is concomitant with the notion of thermodynamic equilibrium (in short equilibrium).
For solids, even to assess which state is equilibrium and which is nonequilibrium is not a trivial matter. For example, today it is a standard view that glass is a nonequilibrium state, even though the state is static by any means. It is not easy to say whether quasicrystals are equilibrium states or not.
Therefore, at the outset, we need to give a definition of equilibrium even in a rough manner: otherwise we could not envisage which kinds of states are going to be studied in this paper. For a moment, let us understand equilibrium as a state that the macroscopic properties of a solid do not change during the time period in question. A rigorous definition of equilibrium is given later. Nobody doubts to regard static states of crystals as equilibrium. However, defect states of crystals are already problem, because defects are the source of hysteresis: the types and concentrations of defects depend on the conditions of crystal growth. In the present viewpoint, these defect states must be taken as the subjects of the study, as far as the defects are immobile. We wish to construct general principles, meaning that no exception should be left merely because of its aperiodicity, exotic properties, whatever the reason is. Glass is, of course, no exception. However, the glass research has a long history, and it has even now many conceptual difficulties. Hence, the topics of glass are presented in separated papers \cite{Shirai20-GlassStates,Shirai21-GlassSHeat}.
Experimentalists often observe hysteresis in solids when changing an external field. Magnetic hysteresis is observed by sweeping an external magnetic field. During changing the field, the state of the solid is admittedly a nonequilibrium state because of time dependence. However, when halting changing the external field and after waiting a sufficiently long time, the state finally reaches a static state. Thus, the final state is an equilibrium state from the present viewpoint, and hence it must be taken as the subject of this study \cite{Note-3}.
However, magnetic properties are not considered here, even though the principles are the same. This is because an additional freedom of spins makes the argument unnecessarily complicated.

The remainder of this paper consists of the following. Section \ref{sec:fundaments} addresses the definitions of thermodynamic equilibrium and the related topics. The background theory is based on the works of Gyftopoulos and Beretta (GB) \cite{Gyftopoulos} and Reiss \cite{Reiss}.
On this basis, we start to study the thermodynamic description of solids in Sec.~{\ref{sec:main}. Many questions associated with the present argument are related to the notion of constraints, and accordingly, this notion is discussed in depth in Sec.~{\ref{sec:constraints}. Some working examples of TCs of solids are given in Sec.~{\ref{sec:examples}. Finally, a summary is given in Sec.~{\ref{sec:conclusion}. The author received common questions from readers from time to time, and hence it may be useful to collect important questions and the answers in Supplemental materials.
Throughout this paper, entropy $S$ is presented in the unit of Boltzmann's constant $k_{\rm B}$, and that, in the product $TS$, $S$ is evaluated with $k_{\rm B}$ is understood.

%%%%%%%%%%%%%%%%%%%%%%%%%%%%%%%%%%%%%%%%%
\section{Fundamental relation of equilibrium}
\label{sec:fundaments}

Let us begin with the definitions of thermodynamic equilibrium and state variables in a general manner. 
% vain  useless  % viable
As stated above, any attempt to define equilibrium by the words that are defined within equilibrium will eventually fail, because the argument is circular. Only a way to escape from this contradiction is to start argument from the general case of nonequilibrium and to treat equilibrium as a special case of nonequilibrium. 
This is the approach that GB employed for constructing the foundation of thermodynamics \cite{Gyftopoulos}, which is a significant improvement of their predecessors of Hatsopoulos and Keenan \cite{Hatsopoulos}.
Since the GB approach is very different from the traditional manner of teaching thermodynamics in terms of both ideas and terminologies, it is difficult to explain their approach in a short passage. 
A brief exposition of the GB approach in the present context is given in Appendix A.
For an in-depth understanding the GB approach, however, readers are encouraged to refer to their original textbook \cite{Gyftopoulos}.

The GB approach begins with very general situations: no assumption of equilibrium is made. Let us consider a system consisting of many particles; by system, it is referred to in this meaning throughout this study. A system in nonequilibrium is characterized by a set of state variables $X(t)$; the state variables differ from the traditional meaning of state variables: any physical quantity determined by well-defined measurements, regardless of whether it is time-dependent (dynamic variable) or not, is a state variable. The number of particles in a small segment of space, $n({\mathbf r},t)$, is an example.
In order to avoid confusion, the word TC is restricted to refer to the state variable in the traditional meaning: it is defined only for equilibrium states. After this section, we treat only equilibrium states, and hence the word state means an equilibrium state, as in the traditional manner of thermodynamics.

The GB approach next defines equilibrium. Suppose that we have an adiabatic system connected to a weight as the sole device external to the system.

\noindent
{\bf Definition 1: Thermodynamic equilibrium}

{\em It is impossible to change the stable equilibrium state of a system to any other state with its sole effect on the environment being a raise of the weight}. 

\noindent
The statement of its sole effect on the environment being a raise of the weight means in usual terms that work can be extract from the system without leaving any effect on the environment. If two systems $A$ and $B$ are in relation of equilibrium, contact of $A$ and $B$ does not cause any change. If these two are not in that relation, the contact causes a spontaneous change: an increase in entropy. The maximum work theorem states that an increase in entropy can be converted to net work on the environment. Hence, equilibrium state means no spontaneous change in the macroscopic properties of a system, which agrees with an intuitive understanding of thermodynamic equilibrium.

By defining equilibrium by Definition 1, we can avoid the circular definition between state variables and equilibrium. Then, a TC can be defined.

\noindent
{\bf Definition 2: Thermodynamic coordinate}

{\em A thermodynamic coordinate is the time-averaged value of a property under a constraint on the range in which the temporal value of the property can vary}. 

\noindent
By defining a TC $X_{j}$ in this manner, it is understood that the time average of the corresponding property $X_{j}(t)$ converges to a finite value, $X_{j} = \bar{X_{j}}$. 
A constraint $\xi_{j}$ is a means of restricting the range in which a particular variable $X_{j}(t)$ can vary. It is expressed by either a hypersurface $\xi_{j}(X_{j}(t))=0$ or a range $\xi_{j}(X_{j}(t))>0$. Although the restriction can be applied to any mode of change, for a moment, restrict ourselves to those in the real space. Later, different types of constraint are shown on occasions. An example of a constraint $\xi$ is the rigid wall of a container in which an enclosed gas can move. This example is easily understandable. On the other hand, in real materials, the notion of constraint seems not clear. The substance of constraint in real materials is an energy barrier, which restricts a range of atom motion within a unit cell or smaller region. In other words, the structure is akin to constraint, which identifies a given type of solid from others.
Further investigation of constraint is given in Sec.~\ref{sec:constraints}.
A constraint determines a TC. The constraint of a rigid wall of a container specifies volume $V$, which is a TC  for the gas inside the container when the gas is in equilibrium. This relationship between a TC and a constraint is well described by Reiss \cite{Reiss} (Sec.~I.8), who may be the first to seriously investigate this relationship, whereas it is not clear in GB. In addition, although in GB two types of constraints, {\it i.e.}, external and internal constraints, are distinguished, the distinction is not important for the present argument, and accordingly in this paper word ``constraint" is used to refer to both types. 

In terms of the equilibrium state, the second law is expressed as follows.

\noindent
{\bf Theorem 1: The second law of thermodynamics}

{\em Among all the states of a system that have a given $U$ and are compatible with the given constraints, there exists one and only one stable equilibrium state.} 

\noindent
A stable equilibrium state is meant as the equilibrium state that can retain its state upon finite perturbations. Hereafter, an equilibrium state is considered to be stable unless otherwise stated. 
This stable equilibrium state corresponds to the maximum entropy state in usual terms.

An important feature of equilibrium is its relationship with TCs, which plays a central role in the present study.

% Corollary
\noindent
{\bf Corollary 1: Existence of the fundamental relation of equilibrium}

{\em Under given constraints with a fixed $U$, the thermodynamic coordinates $\{ X_{j} \}$ of a system are uniquely determined when the system is in equilibrium.}

\noindent
This is a direct consequence of Theorem 1. For each set of constraints $\{ \xi_{j} \}$, there is one and only one equilibrium state, and hence the equilibrium state fixes only one value $X_{j}$ for each $\xi_{j}$. 
Some $X_{i}$ may not be independent of the others. In this case, by restricting a set $\{ X_{j} \}$ to the set containing only independent variables $\{ X_{j} \}_{j=1, \dots, M}$, this uniqueness is warranted. A formal proof is given in Ref.~\cite{Munster-ST}.
The uniqueness of $\{ X_{j} \}$ for a given equilibrium state entails the existence of an equation that relates all the TCs $\{ X_{j} \}$ uniquely. This equation is called {\it the fundamental relation of equilibrium} (FRE) \cite{FundamentalEquation}

\begin{equation}
S=S(U, X_{1}, \dots, X_{M}),
\label{eq:FundamentalE}
\end{equation}
in the entropy representation \cite{Gibbs}. By changing the variables to other sets, the free energy takes the same role.
Corollary 1 guarantees the existence of the FRE, but a concrete form of the equation cannot be obtained from the laws of thermodynamics alone. 
To obtain a concrete form, we need specific models, empirical rules, or help of statistical mechanics. This is true even for an ideal gas, the equation of the internal energy $U=(3/2)RT$ ($R$: the gas constant) is obtained on the assumption that the specific heat is independent of $T$ and $V$. 

A special form of the FRE is useful for analyzing TCs. We know that the FRE of ideal gases is given by $S=\ln (UV/N)$ with an appropriate reference state. As in this example, when the Hamiltonian of a system can be diagonalized with respect to $\{ X_{j} \}$, $S$ is factorized as
\begin{equation}
S= \sum_{j}^{M} s_{j}(X_{j}),
\label{eq:DiagonalS}
\end{equation}
where $s_{j}(X_{j})$ are the respective components. This elegant form is particularly useful when how many TCs are involved is investigated.

The FRE is required to have the following property [3(c), 11, 19]. %\cite{Gibbs, Callen, Gyftopoulos}.

\noindent
{\bf Corollary 2: Completeness of thermodynamic coordinates}

{\em Any thermodynamic property of a system can be obtained from the fundamental relation of equilibrium.}

\noindent
Among the various thermodynamic properties, the most important property may be specific heat $C$. We use this requirement as a guide to obtain the desired set of TCs in the next section.
The requirement that specific heat is obtained from the FRE is a necessary condition---but not sufficient---for Corollary 2 to hold. 

It is very important to recognize that a solid has indeed numerous equilibria for a given $T$ and $V$, whereas a gas has only one equilibrium.
Gyftopoulos and Beretta note that the existence of numerous equilibria combined with the second law of thermodynamics clouds the picture of stable equilibrium, making it difficult to understand (Ref.~\cite{Gyftopoulos}, p.~63). 
We see that, for a solid, a variety of equilibrium states are created by imposing different constraints. By changing the volume, a different state appears. This means that $V$ is a TC. In a similar manner, we can create inhomogeneous strains on a solid. In the hardness measurement, an indenter yields a local strain on a sample. By applying a number of nanoindenters on different places of a solid, we can create an arbitrary shape of deformation (Fig.~\ref{fig:1}). All these deformations are equilibrium states with different constraints. This follows by Definition 1: we cannot extract work from these deformed solids without changing the deformations.
From Corollary 2, we conclude that extra TCs other than $T$ and $V$ are needed. 
On the other hand, for a gas, the deformations of shape, while maintaining a constant $V$, has no consequence on the thermodynamic properties of the gas, and thus the shape is not a TC. 

\begin{figure}[htpb]
\centering
\includegraphics[width=80 mm,bb=0 0 510 255]{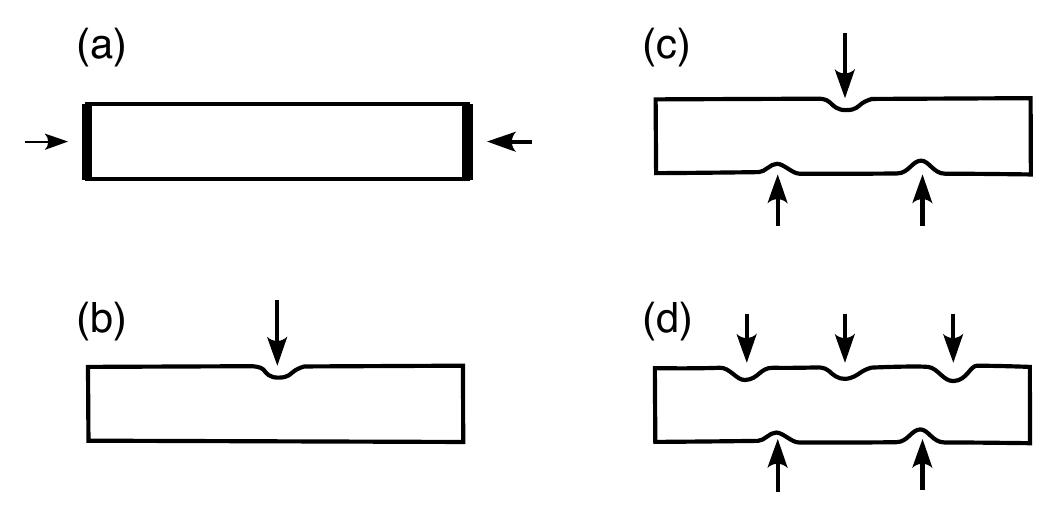}
\caption{Different constraints create different equilibria. (a) A homogeneous deformation has an equilibrium state which is specified by a uniform strain $\varepsilon$. (b)-(d) Local deformations create different equilibria which are specified by local strains $\varepsilon({\mathbf r})$. }
\label{fig:1}
\end{figure}

%%%%%%%%%%%%%%%%%%%%%%%%%%%%%%%%%%%%%%%%%
\section{Thermodynamic coordinates in solids}
\label{sec:main}

We can now construct the thermodynamic description of a solid.
Suppose that a solid is composed of $N_{\rm at}$ atoms, which have their respective positions of equilibrium $\left\{ \bar{\bf R}_{j} \right\} $. We do not assume any crystal symmetry, so that any solid, such as crystals possessing defects and amorphous solids, can be applied to. Only the requirement is that they have equilibrium positions $\left\{ \bar{\bf R}_{j} \right\} $.

Corollary 2 requires that an appropriate set of TCs for a solid must completely describe the specific heat of the solid. It is an elementary task in statistical mechanics to calculate the free energy $F$ of a solid by expanding an atom position ${\bf R}_{j}(t)$ around its equilibrium position $\bar{\bf R}_{j}$ as ${\bf R}_{j}(t) = \bar{\bf R}_{j} +{\bf u}_{j}(t) $ (for example, Ref.~\cite{Reif}, Sec.~10.1). In the harmonic approximation, by transforming atomic displacements $\{ {\bf u}_{j} \}$ to normal modes $\{ q_{k} \}$, the free energy $F$ is obtained as
\begin{equation}
F = k_{\rm B} T \sum_{k} \ln \left[ 2 \sinh \left( \frac{1}{2} \beta \hbar \omega_{k} \right) \right],
\label{eq:FofHarmonicSolid}
\end{equation}
where $\beta$ is the inverse of temperature, $\omega_{k}$ the frequency of $k$th normal mode, and $\hbar$ the Planck's constant. The thermal average $\bar{q}_{k}$ is related to the Bose occupation number $\bar{n}_{k}$ through
% \begin{equation}
$ 
(1/2) m \omega_{k}^{2} \bar{q}_{k}^{2} =\left\{ (1/2) + \bar{n}_{k} \right\} \hbar \omega_{k},
$
% \label{eq:Qtonw}
% \end{equation}
where $m$ is the mass of the constituent atoms, when the crystal is a monatomic crystal.
By using $\bar{n}_{k}$, the entropy $S=-\partial F/\partial T$ of a solid is expressed in the factorized form
\begin{equation}
S = \sum_{k=1}^{M} \left[ (\bar{n}_{k}+1) \ln (\bar{n}_{k}+1) - \bar{n}_{k} \ln \bar{n}_{k} \right],
\label{eq:SofPhonons}
\end{equation}
where $M=3 N_{\rm at}-3$. 
The missing three freedoms together with the symmetry reason give rise to the six components of the macroscopic elastic contribution, $V \sum_{ij} \sigma_{ij}d \epsilon_{ij}$ \cite{BornHuang} (p.~132). This term is not interested here.
From this, the specific heat $C$ is immediately obtained as 
\begin{equation}
C=T \frac{\partial S}{\partial T}.
\label{eq:specfic-heat}
\end{equation}
Here, the $0$th coordinate $U$ is implicitly included in the expression, which has been transformed to another variable $T$. From Eqs.~(\ref{eq:SofPhonons}) and (\ref{eq:specfic-heat}), we have seen that all the coordinates $\{ \bar{n}_{k} \}$ are needed to fully describe the specific heat. The coordinate $\bar{n}_{k}$ is determined by the frequency $\omega_{k}$ of the $k$th normal mode for a given $T$. The set $\{ \omega_{k} \}$ of all the normal modes is determined by the atom potential $\Phi(\{ \bar{\bf R}_{j} \})$ through force constants $\{ \partial^{2} \Phi/ \partial {\bf R}_{i} \partial {\bf R}_{j} \}$ \cite{BornHuang}.

We can displace atoms in a crystal by, for example, electron irradiation. The displaced atom will find its equilibrium position at one of many local minima in the atom potential $\Phi$. This defect state is stable, unless annealed out at a high temperature.  A different atom configuration $\{ \bar{\bf R}_{j}' \}$ alters the phonon spectrum through the change in force constants, and specific heat is changed accordingly. In this manner, we have seen that specific heat of any state of a solid is a function of $\{ \bar{\bf R}_{j} \}$.
Conversely, the set $\{ \bar{\bf R}_{j} \}$ is, by construction, uniquely determined by an equilibrium state, which meets the requirement of Corollary 1. 
Therefore, we conclude that all the averaged atom positions are TCs for a solid. 
Entropy $S$ is obtained by integrating $C$ in Eq.~(\ref{eq:specfic-heat}), and hence entropy retains all TCs $\{ \bar{\bf R}_{j} \}$.

\noindent
{\bf Corollary 3: Thermodynamic coordinates of solids}

{\em The fundamental relation of equilibrium for a solid is represented by the time-averaged position of all atoms that comprise the solid, in addition to $U$.}
\begin{equation}
S=S(U, \bar{\bf R}_{1}, \dots, \bar{\bf R}_{N_{\rm at}}).
\label{eq:FundamentalEofSolid}
\end{equation}

\noindent
The dimension of the state space is $3 N_{\rm at}$, aside from $U$. There are as many equilibrium states as the number of allowed points in this state space. Once entropy was obtained, it is a standard procedure to convert other thermodynamics functions, which are again functions of $\{ \bar{\bf R}_{j} \}$.
% By the arguments up to here, we see that a set of constraints of a material is no more than the structure of the material, because the constraints determine $\{ \bar{\bf R}_{j} \}$. The substances that keep the structure are energy barriers for atom position $\{ \bar{\bf R}_{j} \}$.
%
% Gyftopoulos and Beretta urge that the FRE is derived as a rigorous consequence of the first and second laws, not as a consequence either of difficulties related to exact calculations and lack of knowledge, or a need to describe complicated physical problems by a few gross macroscopic averages (\cite{Gyftopoulos}, p.~119). The present conclusion of Corollary 3 is rigorous to the extent of this strength.
%Ref: Maradudin

Some readers may be averse to accepting this conclusion. The followings are important concerns that have commonly been raised.
One concern is that atom positions are microscopic quantities and hence conflict with the macroscopic nature of thermodynamics. However, we may ask why they are microscopic and are so compared with what. Even our planet is microscopic when the evolution of the universe is investigated. 
In crystal growth, crystal nuclei are small compared with the human scale, but we have elegant thermodynamics theories for crystal growth. In laser cooling, an atom cluster is confined by a photon field \cite{Chu85,Hartmann04}. Although the cluster is microscopic in the human scale, we can measure the temperature of the cluster; otherwise the concept of cooling loses its meaning. An atomic nuclear has its own temperature and even exhibits a phase transition inside it \cite{Povh}. Thermodynamics holds even for extreme matter of black holes \cite{Dvali15}. % although it is not sure whether it is microscopic or macroscopic 
Thermodynamics must hold for phenomena of any scale, and we should not introduce the human scale into a universal theory.

Another concern is that the number of TCs in the above argument is infinite. We learned that a great merit of thermodynamics is the small number of variables used to describe the properties of a macroscopic system [3(c)]. The author agrees with this view for introductory courses of thermodynamics. However, for more complex systems, this restriction of the number of variables no longer makes sense, as is already shown in Introduction. Instead, the important features of TCs consist in Definition 2: distinguishability of a TC from other TCs by constraints and its unique value in equilibrium. There is a double infinity in thermodynamic states of solids: the number of TCs $\{ \bar{\bf R}_{j} \}$ is infinite, while each TC, $\bar{\bf R}_{j}$, consists of an infinite number of the values of a dynamic variable ${\bf R}_{j}(t)$. The latter type makes it sense to enumerate microstates, giving the mean value. The mean value is assigned to the former.
By combining the above two answers, we may better say the nature of thermodynamics: thermodynamics describes equilibrium properties of many-particle systems (macroscopic systems) by a set of (a few number of) state variables (the words in brackets are those of the conventional expression). Accordingly, the conventional word ``macroscopic" property should be replaced with an average property. 

The outcomes of statistical mechanics and thermodynamics must match each other: otherwise, one of the theories (or both) must be wrong. Calculation of the partition function of a solid from its phonon spectrum $g(\varepsilon)$ belongs to tasks of the statistical mechanics, given that $g(\varepsilon)$ is known: the formula Eq.~(\ref{eq:FofHarmonicSolid}) for the free energy $F(T)$ is obtained accordingly. In thermodynamics, $F(T)$ can be obtained from experimental data on specific heat as a function of $T$. 
These two quantities, $F(T)$ and $g(\varepsilon)$, are related by a Laplace transformation,
\begin{equation}
F(\beta) = \int_{0}^{\infty} g(\varepsilon) e^{-\beta \varepsilon}d\varepsilon.
\end{equation}
Two theories of thermodynamics and statistical mechanics are equivalent in this sense \cite{Mandelbrot64,Tisza63}. If the one function is a function of $\{ \bar{\mathbf R}_{j} \}$, then the other function is too.

%%%%%%%%%%%%%%%%%%%%%%%%%%%%%%%%%%%%%%%%%%%%
\section{Roles of constraints and equilibrium properties}
\label{sec:constraints}

\subsection{Creation of thermodynamic coordinates}
% \subsubsection{Role of constraints}
\label{sec:roleConst}

In view of importance of the relationship between constraints and TCs, we further elaborate on the meaning of constraints. The idea of introducing constraints began by Gibbs a century ago, whereas a different name {\it passive resistance} was used \cite{Gibbs}. He already noticed that metastable states are sustained by constraints. 
A constraint $\xi$ restricts the value of a dynamical variable of a system in a range that characterizes the problem under consideration. For a container filled with a gas, the wall $\xi_{1}$ of the container restricts the space that the gas can occupy, defining $V_{1}$ as a TC. When an internal wall $\xi_{2}$ is inserted, a pair of TCs, $V_{1}$ and $V_{2}$, appears. In this way, a constraint creates a TC. Reiss emphasizes this role of constraints: there is a one-to-one correspondence between TCs and constraints \cite{Reiss}. 
The number $M$ of TCs is the same as the number $K$ of constraints $\xi_{j}$.
In the above example, the constraint acts on an extensive quantity $V$. For a crystal, the boundary of a cell $\xi_{j}$ restricts an atom in the cell. Since the cell boundary is a fictitious construct, a better way in this case is that the force ${\bf F}_{j}$ acting on that atom is taken as the constraint. In this case, the constraint acts on an intensive quantity. 

The mechanism of the creation of a TC by a constraint is clearly seen in statistical mechanics. For an $N$-particle system, the potential part of the partition function $Z_{P}$ is expressed as
\begin{equation}
Z_{P} = \frac{1}{N!} \int_{V} \exp[-\beta \phi( {\bf R}_{1}, \dots, {\bf R}_{N} )] d {\bf R}_{1}, \dots, d {\bf R}_{N},
\label{eq:parttition}
\end{equation}
where $\phi$ is an $N$-particle potential. The integration is taken over the entire volume $V$ of the system. For an ideal gas, all the coordinates are integrated out, resulting in $V^{N}$. Only one TC has been created for ideal gases with a fixed $N$. When all the particles are well localized within small spaces $v_{j}$, which do not overlap each other, the multiple integral becomes a product of single integrals as follows:
\begin{equation}
Z_{P} = \frac{1}{N!} \int_{v_{1}} \exp[-\beta \phi_{1}( {\bf R}_{1} )] d {\bf R}_{1} \cdots 
\int_{v_{N}} \exp[-\beta \phi_{N}( {\bf R}_{N} )] d {\bf R}_{N}.
\label{eq:parttition1}
\end{equation}
For each factor, the integration is taken only over the cell $v_{j}$. The instantaneous position ${\mathbf R}_{j}$ is close to its mean position $\bar{\mathbf R}_{j}$, so that it can be expanded as ${\mathbf R}_{j} = \bar{\mathbf R}_{j} + {\mathbf u}_{j}$ by a small  displacement ${\mathbf u}_{j}$.
By integrating with respect to ${\bf R}_{j}$, an averaged coordinate $\bar{\bf R}_{j}$ is singled out from the integration as
\begin{equation}
\int_{v_{1}} \exp[-\beta \phi_{1}( {\bf R}_{1} )] d {\bf R}_{1} =
\exp[-\beta \phi_{1}( \bar{\bf R}_{1} )] \int_{v_{1}} \exp[-\beta \phi_{1}( {\bf u}_{1} )] d {\bf u}_{1}.
\label{eq:parttition1a}
\end{equation}
In this manner, $\bar{\bf R}_{j}$ is created for each factor, leading to Eq.~(\ref{eq:FundamentalEofSolid}). 

For a general case, a TC $X_{j}$ is defined by,
\begin{equation}
X_{j} = \int \xi_{j}( {\bf R}_{1}, \dots, {\bf R}_{s}) d {\bf R}_{1}, \dots, d {\bf R}_{s},
\label{eq:generalX}
\end{equation}
where $\xi_{j}( \{ {\bf R}_{k} \} )$ represents the constraint $\xi_{j}$: a function connecting a set of $\{ {\bf R}_{k} \}_{k=1, \dots, s} $ to a single variable $X_{j}$. The integration is taken only over those coordinates which are allowed to vary within the constraint, and $s$ is the number of those atoms. 
This form is useful for treating atom clusters and domain structures.

The same conclusion can be obtained from information theory, where the principle of maximum entropy plays the central role \cite{Jaynes57,Jaynes78,Grandy}. 
For a given set of discrete values $\{ x_{j} \}$ of a statistical quantity $x$ and a set of $m$ kinds of expectation values $\sum_{j} p_{j} f_{r}(x_{j}) = F_{r}$ for $r=1, \dots , m$, the optimal choice of a probability distribution $\{ p_{j} \}$ is obtained by maximizing the information entropy $S_{I}= -\sum_{j} p_{j} \ln p_{j}$ subject to constraints
\begin{equation}
\sum_{j} p_{j} f_{r}(x_{j}) = F_{r}, \quad for \ r=1, \dots , m
\label{eq:Lagrandians}
\end{equation}
Here, $f$ and $F$ are not forces but given functions and their expectation values, respectively. The standard procedure of solving this problem is use of the method of Lagrange multiplies $\{ \lambda_{j} \}$. 
Thus, we have seen that a TC in thermodynamics is nothing more than an expectation value in information theory. Therefore, it is legitimate that the number of constraints $m$ is equal to the number of TCs $M$. This equality was established by the work of Jaynes for a long time ago \cite{Jaynes57}. However, the view that the number $M$ must be small had so prevailed that he did not touch upon the problem how many TCs exist. A contribution of the present study is clarifying that atom positions can be expectation values for solids.

%%%%%%%%%%%%%%%%%%%%%%%%%%%%%%%%%%%%%%%%%
\subsection{Ergodicity and constraint}
\label{sec:timescale}

% Constraint has an important relation to ergodicity. 
Equilibrium means ergodicity. Ergodic state is understood as the state that every particle in a system visits everywhere in the system with equal probability. However, this equal probability is often misconstrued, when the break of equal probability is brought about by constraints. Penrose noted that, for non-ergodic systems, the energy manifold can be partitioned into invariant submanifolds (\cite{Penrose79} p.~1947). Ergodicity has sense only when constraints are specified. 

Even an ideal gas in a container breaks ergodicity in the sense that the molecules in the container do not visit the space outside the container. Obviously, this is due to the wall of the container inhibiting the molecules from escaping from the container. But, in the real world, no containers are free from gas leak, whatever small. Any vacuum apparatus eventually loses its vacuum pressure if a vacuum pump is not operated. The potential barrier built in metal walls impedes the diffusion of gas toward the outside. The finite value of an energy barrier makes the relaxation time $\tau$ of gas diffusion a finite. A rigid container is an idealized device to simplify physical problems by taking $\tau$ infinite.
Even if the wall is removed, perfect ergodicity is not recovered. Air molecules cannot visit the space outside the atmosphere of the earth. Furthermore, even within the atmosphere, air molecules do not visit spaces of different elevations with equal probability. The break of equal probability is brought about by the gravitational potential $V(\bf r)$. Potential $V(\bf r)$ is not a sort of allow or inhibit. The distribution of particles are continuously changed according to $V(\bf r)$. We should not require equal probability of visit for equilibrium.

Real constraints also impose the validity of equilibrium in a limited time scale.
In the chemical reaction ${\rm H_{2} + (1/2) O_{2} \rightarrow H_{2}O }$, the enthalpy of the product is lower than that of the reactants (by 280 kJ/mol), from which we expect the reaction to take place. 
Despite this, the mixture of hydrogen and oxygen gases is stable at room temperature. The energy barrier between hydrogen and oxygen gases prevents the chemical reaction from taking place. However, the stability of this mixture is broken either by raising temperature or by waiting for long times. 
% When nuclear reactions are taken into account, there is no stable matter. 
The above examples indicate that equilibrium has sense only within a relaxation time. A material consists of many constraints, and each one has its own relaxation time. The relaxation time that controls the equilibrium depends on the scope of problem in question.
This caution is sometimes forgotten for investigating special kinds of materials. Glass is said to be a nonequilibrium state for the reason that, if we wait for a long time, it will crystalize. 
% However, there is no evidence for the crystallization of glasses \cite{Berthier16}. 
Such an argument based on speculation for future events is not a legitimate way to assess the equilibrium of the present state. If the present state is static, the state must be indeed equilibrium. Further discussion on the stability of glass is discussed in another paper \cite{Shirai20-GlassStates}.
% Now that various materials having ambiguous structures, such as incommensurate crystals and quasicrystals, emerged, it is inappropriate to assess the stability of such materials from speculated future events.

%%%%%%%%%%%%%%%%%%%%%%%%%%%%%%%%%%%%%%%%%
\subsection{Frozen coordinates}
\label{sec:frozen}
There is an interesting class of constraints.
When Kline and Koenig investigated the definition of state variable, they used the following example \cite{Kline57}.
The properties of water in a container depend on $T$ and $V$, whereas the elevation $H$ at which the container is held is irrelevant. The FRE is expressed by these two coordinates, $(T, V) \equiv \{ X_{j}^{0}\}$. 
However, there is a case that $H$ affects the properties of water. In a hydraulic power plant, $H$ is the most important coordinate. When the gate of the upper pool is closed, the thermal communication between the top and bottom water is inhibited, and $H$ does not enter the FRE. Here, we refer to this silent coordinate as the {\it frozen coordinate} and denote it by adding a caret, $\hat{H}$. The FRE is then expressed as $S(\{ X_{j}^{0}\}; \hat{H})$. In this case, the role of the constraint is to completely inhibit $\hat{H}$ from varying. When the gate is opened, the top water falls into the bottom pool, $H$ begins to vary and becomes a true variable. This is the process in which the gravitational potential of the top water is dissipated to the thermal energy of the bottom water. Now, we must take $H$ into account in $S$, as $S=S(\{ X_{j}^{0}\}, H)$.
% Isotope, freedom of spins, nuclear spins

The silicon crystals used in the electronics industry are most perfect crystals ever obtained. Despite this, a disorder due to the random distribution of isotopes lurks behind the structural perfection. This random distribution does not contribute to the thermodynamic properties of silicon, and hence the positions of isotopes are frozen coordinates. When an isotope enrichment process is investigated, however, the distribution of isotopes becomes a real variable, and hence we have to take them into consideration. 
In sum, a frozen coordinate $\hat{X}$ is such a coordinate that the energy barrier corresponding to $\hat{X}$ is so high that change in $\hat{X}$ is virtually inhibited within the scope in question.

The above argument shows that the absolute value of entropy makes no sense.
We can add  an arbitrary number of frozen coordinates $\hat{X}_{k}$ without causing any change in the thermodynamic properties of the system,
\begin{equation}
S= \sum_{j}^{M} s_{j}(X_{j}) + \sum_{k}^{ } s_{k}(\hat{X}_{k}).
\label{eq:sumplus}
\end{equation}
% We have freedom to add arbitrary frozen coordinates as we wish. 
This arbitrariness in $S$ was pointed out previously \cite{Grad61}. Sometimes it is called the ``anthropomorphic nature" of entropy \cite{Jaynes65}.
Nonetheless, each component $s_{k}(\hat{X}_{k})$ has its absolute value, which makes it sense to compare $s_{k}(\hat{X}_{k})$ among different materials. This is an issue of the third law of thermodynamics, and is studied in depth in Ref.~\cite{Shirai18-res}.

%%%%%%%%%%%%%%%%%%%%%%%%%%%%%%%%%%%%%%%%%
\subsection{Degeneration of the state space}
\label{sec:degeneration}
A constraint is linked to the notion of missing information (lack of detailed information), which is a useful interpretation of entropy \cite{Haar-Thermostat,Ben-Naim,Jaynes57}. 
Missing information implies an increase in uncertainty. 
It is brought about most drastically when degeneration of the state space occurs, as described below.  Although the number of atoms $N_{\rm at}$ is not altered in a transformation from a solid phase to a gas phase, readers may wonder why the number of TCs changes markedly from the order of $10^{23}$ to only 2. 
The point is how many coordinates are {\em uniquely} determined in equilibrium state. For a gas state, information of detailed positions of atoms is irrelevant to thermodynamic properties of the gas in equilibrium.

For perfect crystals, all the positions of the atoms ${\bf R}_{j}$ are confined in their respective unit cells, whose volume is $v_{c}$. We can experimentally determine $\bar{\bf R}_{j}$ with certainty. 
The entropy per atom can be roughly estimated by $\ln (v_{c}/v_{Q})$, where $v_{Q}=\lambda_{T}^{3}$ and $\lambda_{T}$ is the thermal de Broglie wavelength. Since the size $v_{c}$ is comparable to $v_{Q}$, $S$ is close to zero.
The boundary of a unit cell is a wall of the $j$th constraint $\xi_{j}$. Let us remove the first wall $\xi_{1}$. Then, atoms 1 and 2 can interchange their positions. It is uncertain which atom is in which cell, except that the two atoms are located somewhere in a volume of $2 v_{c}$. $S$ is increased to $\ln (2 v_{c}/v_{Q})$. We lose either one of two positions from the set of TCs $\{ \bar{\mathbf{R}}_{j} \}$. 
Next, let us remove the second wall $\xi_{2}$. The coordinate of atom 3 is lost in the set $\{ \bar{\mathbf{R}}_{j} \}$, associating with an increase $S$ to $\ln (3 v_{c}/v_{Q})$. In this way, when all the walls are removed, we lose all the positions of atoms from the set of TCs, leaving only the whole volume $V = N_{\rm at} v_{c}$ as useful information. The final state is a gas state, whose entropy is given by $V$ only, as $\ln (V/v_{Q})$.
During this process, the number of TCs has been reduced to one from $3 N_{\rm at}$,
\begin{equation}
( \bar{\mathbf R}_{1}, \dots, \bar{\mathbf R}_{N_{\rm at}}) \; \rightarrow V.
\label{eq:ri-ni}
\end{equation}
It is said that the {\it degeneration of the dimensionality of state space} occurs.
% \footnote{Word degeneration here is used in a different meaning from the usage in Bose-Einstein and Fermi-Direc statistics. A similar usage as the present one is found in Ref.~\cite{Tisza63}.}

Degeneration of the state space can occur, when a system undergoes a phase transition. We can look upon the order parameters in a phase transition as TCs. Anderson figured out that the order parameters in crystallization are atom positions, although he refers to the density in the reciprocal space $\{ \rho(\mathbf G) \}$ for a reason uninterested in the present context \cite{Anderson84} (p.~39). 
% quantify, choose, regards
There is an intimate relationship between a TC and an internal variable in nonequilibrium thermodynamics \cite{Coleman67,Maugin94,Gujrati10}, although the latter is mainly used for describing processes. % This topics is beyond the present scope.

%%%%%%%%%%%%%%%%%%%%%%%%%%%%%%%%%%%%%%%%%
% Multifaceted equilibrium properties
\subsection{Hysteresis and irreversibility}
\label{sec:solids}
At this stage, it is pertinent to discuss the implications of the TCs on solid-state phenomena. 
Corollary 3 holds for any state of solids in equilibrium, irrespective of the internal structure and periodicity.
The fact that a solid exhibits various thermodynamic properties has its origin in the large number of TCs, i.e., $3 N_{\rm at}$, while for a gas information about atom positions has no consequence on the thermodynamic properties of the gas.
We have already seen such an example in arbitrarily shaped deformations in the last paragraph of Sec.~\ref{sec:main}. As the scale of the deformation becomes finer, the deformation eventually becomes atomic displacements. Mechanical properties of metals are altered by heat treatment, and these changes are brought about by the changes in the morphology of grains, dislocations, and any structural defects. 
If we carefully observe the crystal growth of silicon, which is the best-controlled crystal, we will find that the rate of crystal growth alters the concentration of intrinsic defects, resulting in different electronic properties of the obtained wafers \cite{Voronkov82}. The martensite transition, which is used in shape-memory materials \cite{Otsuka02}, is caused by atom displacements forming twin structures. The compound GST exhibits a phase transition between the crystalline and amorphous structures, which is utilized for digital versatile discs (DVDs)\cite{Yamada87}. These changes are observed as hysteresis phenomena, and are traditionally considered as nonequilibrium states.

The presence of hysteresis complicates the distinction between state and process in solids, which obscures the thermodynamic nature of the states of solids. 
There are presently several approaches to study hysteresis from nonequilibrium thermodynamics \cite{Truesdell84,Maugin93,Bertotti05,Jou10}. It is correct to treat hysteresis as nonequilibrium phenomena, when the aspect of process of hysteresis is focussed, as noted in \cite{Note-3}. 
However, when the end state of a process is static, we can describe it by equilibrium thermodynamics.
In the literature, it is often proclaimed that a plastic-deformation material is a nonequilibrium state merely because the process of plastic deformation is irreversible. 
However, this argument mistakes state for process, as already noted in Introduction. The free expansion of a gas in a cylinder is an irreversible process, whereas the final state of the gas in this process is an equilibrium state, which can be specified solely by $V$ and $T$. 
In a similar manner, even though a plastic deformation is an irreversible process, if changing of the applied stress is halted, the plastic motion finally ceases. The final state must be an equilibrium state, because it is completely specified by the present values of $ \{ \bar{\bf R}_{j} \} $. If two samples with exactly the same structures are obtained by different paths of hysteresis, we cannot say which sample is obtained by which path merely by inspecting the present properties of the samples.

Hysteresis in solids arises from structural relaxation, that is, retardation in atom rearrangements against the change of an external field. The relaxation part of the structural change is a spontaneous change and accompanies entropy production. Thus, the process is irreversible. The amount of entropy production depends on the process. This process dependence of the irreversible part of the entropy change $d_{i} S$ is compensated for by the process dependence of the reversible part of entropy $d_{e}S$, recovering the process-independent property of the total entropy of a system, $d S = d_{e} S + d_{i} S$ \cite{Prigogine67}.

There are two kinds of the sources for the irreversible part of entropy change $d_{i} S$. 
The one source is the heat generation due to friction. The other one is the occurrence of degeneration of the state space, as discussed in Sec.~\ref{sec:degeneration}. In almost all cases, structural changes in hysteresis accompany creation of disorders. Different configurations of disorders often have the same energy. There are so many atom configurations of disorders with the same energy that the final state is unpredictable. This unpredictability for the final state---which state among many reachable states is realized---is reflected in the increase in the configuration entropy $\Delta S_{\rm conf}$. Hence, in stating more accurately, the uniqueness of state is guaranteed within the uncertainty due to $\Delta S_{\rm conf}$.
The TCs in this case are $\{ \bar{\bf R}_{j} \}_{j=1\dots N'}$ plus $\Delta S_{\rm conf}$, where $N'$ is the number of atoms that did not change their equilibrium positions during the process. A detailed description of the relations between hysteresis and irreversibility is given in Refs.~\cite{Shirai20-GlassStates,Shirai21-GlassSHeat}.
% This secures the validity of the thermodynamic state for solids from breaking down.

Unfortunately, we do not know a concrete functional form of the FRE, $S(U, \{ \bar{\bf R}_{j} \})$, for solids. The present theory is a theory about existence: it certifies the existence of $S(U, \{ \bar{\bf R}_{j} \})$. Today, it is a common task to calculate the electronic properties of solids by first-principles methods, which are based on the principle that the present properties of a solid are determined solely by the present structure $\{ \bar{\bf R}_{j} \}$. We calculate the memory function of phase-change materials without knowledge of their past history \cite{Massobrio15}.
Fortunately, usually not all the atom positions are necessary for describing a specific property because of the degeneration of TCs. % as described in Sec.~{\ref{sec:degeneration}}. 
How to wisely choose a minimum yet sufficient set of TCs is of practical importance. The present study does not give an answer to this question for individual cases, but provides a theoretical basis ensuring that there must always exist a set of TCs for describing any thermodynamic property of a solid. If it is unclear how to choose a small number of TCs, our suggestion is to use a full set of $\{ \bar{\bf R}_{j} \}$ by sampling real materials, which makes first-principles calculations feasible. Then, find a few atom positions which have crucial importance for the property in question. Study of DVD materials by first-principles calculations \cite{Massobrio15} belongs to this approach. Alternatively, we can start from a small number of traditional TCs and improve the accuracy by introducing further TCs. For plastic deformations of solids, Bridgman used strain and stress as two independent variables \cite{Bridgman50}, otherwise dependent each other in elastic theory. Then, more variables, such as abstract variables called internal variables \cite{Kestin70,Rice75} or concrete variables of the density tensors of dislocations \cite{Berdichevsky06}, are introduced in order to better describe the structural sensitivity of mechanical properties.

%%%%%%%%%%%%%%%%%%%%%%%%%%%%%%%%%%%%%%%%%
\section{Examples of thermodynamic coordinates}
\label{sec:examples}

We have seen that the FRE for solids is fully expressed in terms of the time-averaged positions of the atoms, $\left\{ \bar{\bf R}_{j} \right\}$. %, as Eq.~(\ref{eq:FundamentalEofSolid}). 
However, the idea is so novel that it is desirable to work out more specific examples of TCs in solids. 
For perfect and unstrained crystals, considerable economy is obtained in constructing a set of TCs by expressing $\mathbf X = (\mathbf{abc}; \{ \bar{\bf R}_{\nu}^{\kappa} \}_{\nu=1, \cdots s} )$, where $\mathbf{a},\mathbf{b}$, and $\mathbf{c}$ are lattice vectors, $\mathbf{R}_{\nu}$ are the positions of the basis atoms in a cell, $s$ is the number of basis atoms, and $\kappa$ denotes atom species. 
Here, the convention that Greek subscripts refer to the index of the basis atoms, while Arabic subscripts to the index of all the atoms in a solid, is employed. In the following, $(\mathbf{abc})$ (often $U$ too) are not interested and hence omitted from the coordinates. Therefore, the dimension of the state space is $3 N_{\rm at}$.

\subsection{Defects}
\label{sec:defects}

Let us first consider the defect problem of monatomic crystals. Electronics-grade Si crystal is the most perfect crystal. In this case, the TCs are the equilibrium positions of atoms $\{ \bar{\bf R}_{\nu}^{0} \}$. 
The corresponding entropy is $S_{0}$, which may be zero as $T \rightarrow 0$. At a low temperature, defects can be introduced by electron irradiation. For the sake of simplicity, let us ignore the change in $U$ in this process, and accordingly the phonon spectrum is not altered. Suppose that $n_{I}$ interstitial atoms were created. The variable $n_{I}$ can be controlled by the amount of electron irradiation, irrespective of $T$. These defect atoms are located at $\{ {\bf R}_{j'}^{I} \}_{j'=1, \cdots, n_{I}}$. Hence, the atom coordinates are fully specified by $( \{ {\bf R}_{j}^{0} \}; \{ {\bf R}_{j'}^{I} \})$, where $j$ runs from 1 to $N_{\rm at}-n_{I}$.
In equilibrium, part of the regular positions $\{ {\bf R}_{j}^{0} \}$ gives the set of TCs $\{ \bar{\bf R}_{j}^{0} \}$ and the resultant entropy $S_{0}$ is given by Eq.~(\ref{eq:SofPhonons}).

The problem is part of the defect positions $\{ {\bf R}_{j}^{I} \}$. The time-averaged positions $\{ \bar {\bf R}_{j}^{I} \}$ of the interstitial atoms are also TCs, because of the constraint built around $j$-th atom, fixing the value $\bar{\bf R}_{j}^{I}$. The dimension of the state space of {\em a particular sample} does not change from $3N_{\rm at}$. Since the phonon spectrum does not change, we cannot find any difference in the thermodynamic properties from the perfect crystal. In this sense, $\{ {\bf R}_{j}^{I} \}$ can be regarded as frozen TCs, that is, $\{ \hat{\bf R}_{j}^{I} \}$, and the state space is expressed by  $( \{ {\bf R}_{j}^{0} \}; \{ \hat{\bf R}_{j'}^{I} \})$.
However, there are a large number of ways in which the atoms can occupy interstitial sites. 
Suppose that there are $N_{I}$ interstitial sites in a Si crystal. Among them, $n_{I}$ sites are occupied by atoms, leaving the remaining sites empty. There are $W(n_{I}) = \left( \begin{array}{c} N_{I} \\ n_{I} \end{array} \right)$ configurations of interstitial atoms. This gives the configuration entropy $S_{\rm conf} = \ln W(n_{I})$. Since the internal energy does not change, all configurations give the same properties: the configuration entropy is the sole effect of the interstitial atoms to the thermodynamic properties of the silicon. Accordingly, detailed positions $\{ \bar{\bf R}_{j}^{I} \}$ are missing from the set of TCs, leaving only $n_{I}$ as the TC for interstitial atoms.
This is said that degeneration of the state space occurs as
\begin{equation}
( {\mathbf R}_{1}^{I}, \dots, {\mathbf R}_{n_{I}}^{I}) \; \rightarrow n_{I}.
\label{eq:ri-ni1}
\end{equation}
A commensurate set of the TCs turns to be $\left( \{ \bar{\bf R}_{j}^{0} \}; {n}_{I} \right)$: the dimension of the state space being $3 (N_{\rm at}-n_{I})+1$. The origin of degeneration in this case is different from the gas case described in Sec.~\ref{sec:degeneration}. In fact, the constraints are not removed for the defect case. Indistinguishability between defect configurations is the origin for the degeneration.

At low temperatures, the interstitial atoms are immobile, and hence $n_{I}$ has no temperature dependence. Any quantity that has no temperature dependence is no effect on thermodynamic properties. As a consequence, $n_{I}$ is a frozen coordinate: the state space is expressed by $\left( \{ \bar{\bf R}_{j}^{0} \}; \hat{n}_{I} \right)$. However, on annealing at a high temperature, the constraints on the interstitial atoms are removed, and $n_{I}$ begins to vary: $n_{I}$ becomes a real TC. Provided a sufficient time is given, $n_{I}$ approaches the equilibrium value $n_{I} = N_{I} \exp(-E_{f}/k_{\rm B}T)$, where $E_{f}$ is the formation energy of the interstitial atom.
When we compare entropy between the perfect crystal and defective crystals, we are tacitly assuming that the frozen TCs are activated, because comparison is possible only on the state spaces of the same dimension. Experimentally, the change in entropy can be measured by a reversible change between the two states. For example, a defective crystal can be changed to the perfect crystal through melting. Even though the starting sample has only one configuration of defects, this sample can have thermal communications with samples with other configurations. Hence, the measured entropy difference automatically has the contribution of configuration entropy.

%%%%%%%%%%%%%%%%%%%%%%%%%%%%%%%%%%%%%%%%%
\subsection{Compounds versus random alloys}
\label{sec:compounds}

When atom species are taken into account, an additional label $\kappa$ is needed to identify the constituent atoms, as $ \bar{\bf R}_{\nu}^{\kappa} $. For compounds, each site $\nu$ has the respective species $\kappa$, {\it i.e.}, $\kappa = \kappa(\nu)$, thus the label $\kappa$ is redundant for TCs. % A set of $s$ coordinates $\left\{ \bar{\bf R}_{\nu} \right\}$ suffices to describe the thermodynamic properties of a crystal. 
In contrast, for random alloys, there is no correlation between $\kappa$ and $\nu$; hence a full set of notations $\left\{ \bar{\bf R}_{j}^{\kappa} \right\}$ is needed to identify the atom configuration.
By rearranging $\left\{ \bar{\bf R}_{j}^{\kappa} \right\}$ as the composite expression $\left( \left\{ \bar{\bf R}_{\nu} \right\}; \left\{ \kappa_{j} \right\} \right)$, we can separately consider two degrees of freedom of position and species. The part of position $\left\{ \bar{\bf R}_{\nu} \right\}$ is not a problem; we already know that $S_{0}=S_{0}(\left\{ \bar{\bf R}_{\nu}^{0} \right\})$. The part of atom species $\left\{ \kappa_{j} \right\}$ requires consideration.

Let us consider a random alloy $A_{1-x}B_{x}$. There are $N_{A}=N(1-x)$ $A$ atoms and $N_{B}=Nx$ $B$ atoms. To isolate the effect of mixing from other uninterested effects, we assume the ideal situation that there is no difference in bond energy between $A$ and $B$ atoms. Any atom configuration $c$, therefore, has the same energy. %, $U_{c}=U_{0}$.
The thermodynamic properties of the alloy do not depend on the specific configuration $\{ \kappa_{j} \}$. The information about $\{ \kappa_{j} \}$ is lost from the list of TCs, leaving only the fraction $x$ of $B$ as a relevant TC.
Degeneration of the state space occurs as
\begin{equation}
(\kappa_{1}, \dots, \kappa_{N}) \; \rightarrow x.
\label{eq:rxtox}
\end{equation}
There are $W(x) = \left( \begin{array}{c} N \\ N_{B} \end{array} \right)$ possible atom configurations, and hence the contribution of the atom species to the entropy $S_{\rm conf}$ is given by
\begin{equation}
S_{\rm conf} = \ln W= -N \left[ x \ln x + (1-x) \ln (1-x) \right].
\label{eq:Sconf0alloy}
\end{equation}
When there is no need to compare the entropy of the random alloy to those of other structures, such as liquid, this TC $x$ can be regarded as a frozen variable.

An interesting question is why the atom species $\kappa$ becomes a TC, because $\kappa$ does not alter $U$ and has no contribution in Eq.~(\ref{eq:first-law}).
This problem is similar to Gibbs' paradox of mixing. If mixing is carried out in a reversible manner---remember that this is possible by using semipermeable membranes---the mixing entropy $S_{\rm conf}$ is compensated for by the work ${\mathscr W}$ required for mixing, as $TS_{\rm conf} = {\mathscr W}$. Therefore, the atom species represents an extensive variable via $x$, and the corresponding intensive variable is the diffusion force $-k_{\rm B}T \ln x$.

%%%%%%%%%%%%%%%%%%%%%%%%%%%%%%%%%%%%%%%%%
% \subsection{Configuration entropy}
\subsection{Distinguishability and indistinguishability of glass solids}
\label{sec:configuration}

Treating a specific material is not the focus of this study. However, on account of the generality of Corollary 3, it is desirable to demonstrate usefulness of the present theory for glass, for which there are many difficult problems. 
Currently, it is the standard view that glass is a nonequilibrium state. A reason for this view is that a glass exhibits different properties depending on the process by which it was obtained. Different methods of synthesis produce glasses having different mechanical properties, despite being the same glass \cite{Note-2}. However, if the term ``glass state" is interpreted as referring to a thermodynamic state, this contradicts thermodynamics, as is mentioned in Introduction. Researchers dispose this contradiction by regarding glass states as nonequilibrium states. 
% In addition, it has long been the pervasive view that if we wait for a long time, a glass will crystallize. 
% We call this {\it the time-scale issue}.
However, we have seen in Sec.~\ref{sec:solids} that the process dependence on the obtained properties is a different matter from the principle of thermodynamics that the current properties are determined solely by the current values of TCs. 
% For the second point, we have seen in Sec.~\ref{sec:timescale} that no equilibrium state exists without the restriction of relaxation time.
Definition 1 always gives an unambiguous criterion for equilibrium. If a glass state were nonequilibrium, net work could be extracted by cooling without leaving any effect on the environment. This violates the second law of thermodynamics. Further details are given in recent papers \cite{Shirai20-GlassStates,Shirai21-GlassSHeat}. Here, only a general relationship of the glass state with the TC is mentioned below.

\begin{figure}[htpb]
\centering
\includegraphics[width=10.0 cm, bb=0 0 662 404]{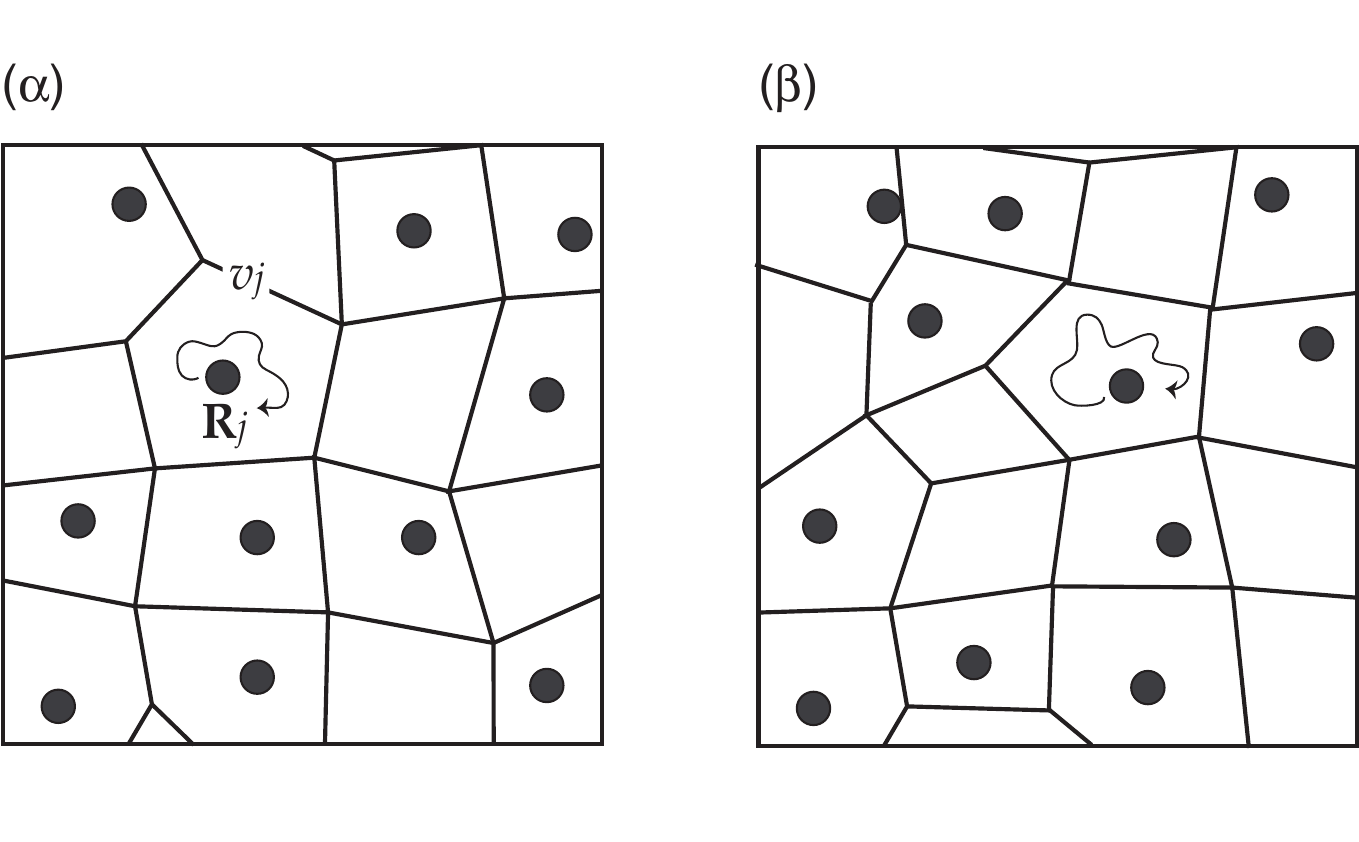}
\caption{Different atom arrangements of a glass. Each atom is constrained by a cell with an irregular shape. For simplicity, the trajectory of motion is drawn for only one atom.}
\label{fig:tiles}
\end{figure}

A glass has many different atom arrangements. Some of them are shown in Fig.~\ref{fig:tiles}. 
The constraint $\xi_{j}$ on the atom at $\mathbf{ R}_{j}$ is its cell boundary $v_{j}$. Each atom can be mobile only within its cell: the exchange of atom positions between different cells is prohibited. One sample of a glass has only one atom arrangement, and another one has a different arrangement. Hereafter, we call a specific arrangement a {\em structure}, while the species of glasses are distinguished by the chemical constitutions and the building units. Let $\alpha$ be the label for a structure and $\{ \xi_{j}^{(\alpha)} \}$ be a set of constraints for the $\alpha$th structure. Both structures $\alpha$ and $\beta$ in Fig.~\ref{fig:tiles} are stable equilibrium states, because they do not alter spontaneously their structures. 
A question is whether these two structures should be regarded thermodynamically as the same state or not. In the present context, they must be different states. This is because, if the two structures were to refer to the same state, it would contradict the second law (Theorem 1) averring that only one equilibrium state is allowed for a given set of constraints. The distinction between different structures is made possible by fully specifying the atom positions. 
Here is a great advantage of employing the statement of Corollary 3: different mechanical properties of a glass are ascribed to different states of the glass. 

A more common case of glass is the indistinguishability of individual samples. A glass does not change its properties from sample to sample if the samples are prepared under the same conditions. 
In experiments, no effect of heat treatment has been observed on the low-temperature specific heat of glasses under a wide range of experimental conditions \cite{Oblad37,Pohl81}. % (p.~31).
The physical and chemical properties of glasses are reproducible; otherwise, they could not be used in high-technology applications.
In this case, the partition function, Eq.~(\ref{eq:parttition}), must contain all the structures that the glass can take from sample to sample,
\begin{equation}
Z_{P} = \sum_{\alpha}^{N_{c}} \frac{1}{N_{\rm at}!} \int_{V} \exp[-\beta \phi^{(\alpha)}( {\bf R}_{1}, \dots, {\bf R}_{N_{\rm at}} )] d {\bf R}_{1}, \dots, d {\bf R}_{N_{\rm at}},
\label{eq:ZofConfif}
\end{equation} 
where $\phi^{(\alpha)}$ is the potential of the ${\alpha}$th structure and $N_{c}$ is the number of different structures that the glass can take. 
Summing over $\alpha$ sometimes causes confusion related to the ergodic problem. 
It is reasonably expected that the integration with respect to one particular structure $\{ {\mathbf R}_{j}^{(\alpha)} \} $ will give the same value among different structures, because the properties of the glass do not change from sample to sample. Accordingly, we have for the entropy of this glass,
\begin{equation}
S=S_{0}(U, \bar{\bf R}_{1}, \dots, \bar{\bf R}_{N_{\rm at}}) + \ln N_{c}.
\label{eq:SofConfif}
\end{equation}
The first term $S_{0}$ on the right-hand side is the phonon contribution from one particular structure and is the same as Eq.~(\ref{eq:SofPhonons}). The last term $S_{\rm conf} = \ln N_{c}$ is the configuration entropy, which  indicates how many equivalent structures exist. The information that is missing is the distinguishability of samples among different configurations $\alpha = 1, \dots, N_{c}$. 
This missing information reduces the dimension of the state space to $N_{\rm at}$ from $N_{c} \times N_{\rm at}$. The extra TC, $N_{c}$, is a frozen coordinate, unless the glass melts. Hence, we can ignore this coordinate, when only properties of a solid state of glass are discussed.
When we compare the entropy of glass to that of the liquid, the TC $N_{c}$ is activated, and thus it should be taken into account. To compare the entropy of glass to that of the crystal, again $N_{c}$ should be taken into account, because the reversible transition between the glass and crystal is possible via melting, while entropy can be evaluated only through a reversible path.

% In the literature of glass physics, another parameter has sometimes been introduced to describe the behavior of glasses \cite{comment4}. This parameter is called the order parameter $Z$ (sometimes called the internal variable). By using $Z$, the enthalpy $H$ is expressed as a function of three variables, $H=H(T,V,Z)$. One example of an order parameter is the fictive temperature $T^{*}$ \cite{Davies53,Moynihan74}. Many order parameters have been devised for a given property \cite{Moynihan81}. A question that has long puzzled researchers is what is the nature of the order parameter $Z$ \cite{Berthier11}. Most researchers deem it as a temporal parameter describing the process and believe that $Z$ vanishes after the completion of the transition. % (see the paper by Gujrati in Ref.~\cite{comment4}). This means that $Z$ is not a TC. From the viewpoint of the present study, if suitably interpreted, order parameters can be looked upon as TCs. A recent study shows that the nature of $Z$ is the degree of overlap of atoms \cite{Charbonneau14}, which is consistent with the present view that the short-range order in the structure determines the thermodynamic properties of glasses. 

%%%%%%%%%%%%%%%%%%%%%%%%%%%%%%%%%%%%%%%%%
\section{Summary}  % and perspective
\label{sec:conclusion}
A solid exhibits various properties depending on the external conditions, such as inhomogeneous deformation, thermal treatment, growth rate, and electron irradiation, other than $T$ and $V$. When the strength of these external influences exceeds a certain level, the solid exhibits hysteresis and its properties are no longer determined by $T$ and $V$ only. How these states are described within the thermodynamic context has been a long-standing problem of thermodynamics. The essential problem is which are TCs that specifies a state of a solid. The present study has given the answer to this question on the basis of the rigorous argument of thermodynamics. The TCs of a solid are the time-averaged positions $\{ \bar{\bf R}_{j} \}$ of the constituent atoms together with the internal energy. The essential principles for arriving at this conclusion are the distinguishability of atom positions and the uniqueness of their values in equilibrium; the former is a consequence of constraints and the latter is a consequence of the second law of thermodynamics. Both properties are lacking for gas states: this is why the properties of gases are quit featureless, while those of solids are structure sensitive.
Corollary 3 guarantees that the present thermodynamic properties of a solid are determined solely by the present state, irrespective of the past history. This secures the foundations of thermodynamics from breaking down for solids.

It is a difficult task to find a FRE expressed by TCs of dimension $3 N_{\rm at}$.
In many cases, fortunately, the full dimension $3 N_{\rm at}$  of the state space is not necessary because of degeneration of the state space. The next question is how to find a minimum and sufficient set of TCs for describing a specific property. Hints are already present in the literature. In glass physics, the properties of a glass are often described by adding only one order parameter to a set of $T$ and $V$. The order parameter is chosen so as to fit a particular problem. This parameter can be a TC. 
% However, careful investigation is needed, depending on the individual problem, as to whether the parameter qualifies as a TC.
The present study provides the theoretical basis that TCs exist for any thermodynamic property of solids and that the effort of finding desirable TCs, in principle, will be successful.

\section*{Acknowledgment}
The author thanks F.~Belgiorno and P. D. Gujrati for valuable debates on various aspects of thermodynamics.
He also thanks Myu Research (www.myu-inc.jp) for the English language review. 
He received financial support from the Research Program of ``Five-star Alliance" in ``NJRC Mater.~\& Dev."

% \clearpage

\appendix

\section{GB thermodynamics}
\label{sec:GB}

In the GB approach, the second law of thermodynamics precedes the zeroth and first laws. It begins with treating general thermodynamic situations: systems are not limited to single-component systems; no assumption of equilibrium is made. The whole of a power plant, ecosystems, even the earth are subjects of study. The essential point of the GB approach is based on the common observation that for any system, whatever complex, the change of state of the system eventually vanishes when the system is isolated from the environment: this is a native expression of the second law. The final state is unique, which is called equilibrium state. For a given system in an arbitrary state, we can adiabatically perform work ($W<0$) on it or can extract work ($W>0$) from it. For a water in an adiabatic container, we can perform work by, for example, stirring it. Conversely, when the water has a convection flow we can extract work from the flow by using, for example, a wheel paddle. While performing work on the system is always possible, there is a special case in which extracting work from the system is impossible. This is the case when the system is in equilibrium. Once the system reaches an equilibrium state, we cannot extract work any more without leaving any effect on the environment. This is generalization of a well-known statement of the second law: it is impossible to obtain work merely by cooling single heat source without leaving any change on the environment. Heat source is a representative of equilibrium system, which is characterized by a single variable, namely, temperature $T$. Existence of $T$ is premised on equilibrium, which is a result of the zeroth law of thermodynamics. By reverting this logical sequence, we can reach Definition 1: an equilibrium state is such a state from which no work can be obtained without leaving any effect on the environment. 
In this manner, we can assess whether the system is in equilibrium or not merely by inspecting the external effects. No information about the internal structure of a system is needed for the definition of equilibrium, as it should be. If the details of internal structure were required, the criterion for equilibrium would alter as our knowledge on the material is developed.

Definition 1 is subject to the restriction that the state of a system is allowed to change only within a given set of constraints. By constraints, it means any barrier prohibiting some freedom of motions, i.e., wall of a container, semi-permeable film separating chemical species, external fields, etc. For a gas in a container of volume $V$, the wall of the container is a constraint, which defines volume $V$ as a TC. If an additional wall is inserted in the container, this creates two volumes $V_{1}$ and $V_{2}$. 
Suppose a chemical reaction $A_{1}+A_{2} \rightarrow A_{3} $, and this reaction does not occur at room temperature. Then, the numbers of these chemical species, $N_{1}$, $N_{2}$, and $N_{3}$, can be taken by arbitrary amounts. What prevents three species from reaction is the energy barrier for the chemical reaction. When this constraint is removed, for example by heating, one of three variables $N_{i}$, cannot be taken as an independent variable. Missing one constraint corresponds to reducing the number of TCs by one. For solids, any kind of energy barrier which prohibits atom motion can be a constraint. 

By using two terms, equilibrium and constraints, the second law is expressed as: {\it among all the states of a system that have a given $U$ and are compatible with the given constraints, there exists one and only one stable equilibrium state} (Theorem 1).
The existence of an equilibrium state is restated, in usual terms, as the state is developed in a manner to achieve the maximum in entropy.  Theorem 1 guarantees the existence of the maximum-entropy state.
But Theorem 1 states more. The existence of {\it only} one equilibrium state is emphasized. From this, the uniqueness of a set of TCs, $\left\{ X_{j} \right\}$ for a given equilibrium state is guaranteed. In this manner, each equilibrium state must be uniquely specified by a set of $\left\{ X_{j} \right\}$. This gives a unique function of $\left\{ X_{j} \right\}$, {\it i.e.}, the fundamental relation for equilibrium.
Impotence of obtaining work from an equilibrium state implies that nonequilibrium states are inevitable for obtaining work. Nonequilibrium is a source of work. When two substances with high and low temperatures are contacted by removing the constraint which isolated them one another, these two substances turn to be a nonequilibrium state. This causes an irreversible change towards the final equilibrium state. Work can be obtained from the thermal flow from the high-$T$ substance to the low-$T$ substance. If this thermal flow is controlled by a reversible machine, the obtained amount of work is maximized. 
There are as many maximum-entropy states as the number of constraints. 
Concrete examples of the existence of variety equilibria for a solid are not given in GB and hence are given here. For silicon wafers, a number of atoms can be displaced into interstitial sites by electron irradiation. Every defect state corresponds a different equilibrium state. Different samples of a glass have different atom arrangements. Each structure of a glass sample is not spontaneously transformed to a different structure at temperatures lower than the glass-transition temperature. Hence, each structure of the glass is a stable equilibrium state. Otherwise, we could obtain work from a glass merely by cooling it.
% When displacements of atoms from their equilibrium positions take place, we can obtain work from the system. When a crystal undergoes a dielectric transition, which is a consequence of displacement of ions, we can obtain electric work.

A notable advantage of the GB approach is that entropy can be defined even for nonequilibrium states. Entropy of a state is defined by the degree how far the state is deviated from the equilibrium state in the given constraints. This degree is quantified by the maximum work that can be obtained within the given constraints. The maximum work is an observable, and hence can be determined by experiment, irrespective of states.
The maximum work for an isolated system is called adiabatic availability. Available energy is an extension of adiabatic availability when the system is in contact with heat sources. They show that available energy is attributed to any arbitrary state of a system, whether it is equilibrium or nonequilibrium. Accordingly, entropy of any state of a system is determined by experiment, whose value is unique to that state.
We do not need to worry about how to count microscopic states for nonequilibrium states, which has been long debated since Boltzmann gave a microscopic definition for entropy \cite{Jaynes65,Lebowitz93}.

%%%%%%%%%%%%%%%%%%%%%%%%%%%%%%%%%%%%%%%%%%%%%%%%%%%
% \bibliography{thermo-refs,glass-refs}

\begin{thebibliography}{54}
\bibitem[{Not({\natexlab{a}})}]{Note-anisotropy}
\bibinfo{note}{For solids, single variable $V$ is replaced by six components of
  strain $\varepsilon_{ij}$. However, these variables are merely extension of
  $V$ to anisotropic media, and have no importance for the present argumentb.}
\bibitem[{Not({\natexlab{b}})}]{Note-3}
\bibinfo{note}{There are a few theories to treat hysteresis based on
  nonequilibrium thermodynamics. Extended irreversible thermodynamics takes
  fluxes into account as fundamental variables \cite{Jou10}. Internal variable
  theory introduce additional freedoms of internal variables into theory
  \cite{Maugin93, Maugin94}. The most radical theory is rational thermodynamics
  \cite{Truesdell84}, viewing thermodynamic phenomena by a part of dynamics.
  All these theories study hysteresis by regarding it as the response to the
  previous influence. This view is correct, if we are interested in how a given material
  has gotten its present structure. 
  However, the present issue is different. The issue is how the present state of the matter
  can be described, when the change of properties of the matter was finished.
  }
\bibitem[{Fun()}]{FundamentalEquation}
\bibinfo{note}{In the literature, this is called the {\it fundamental
  equation}. This name was used by Gibbs (\cite{Gibbs}, p.~85) in old days when there was
  no knowledge of atoms. Today, this name is peculiar and even confusing.}
\bibitem[{not()}]{note1}
\bibinfo{note}{Word degeneration here is used in a different meaning from the
  usage in Bose-Einstein and Fermi-Direc statistics. A similar usage as the
  present one is found in Ref.~\cite{Tisza63}.}
\end{thebibliography}

% \end{thebibliography}

\end{document}